\documentclass[%
 reprint,
groupedaddress,
showpacs,
 amsmath,amssymb,
 aip,
]{revtex4-1}

\usepackage[final]{graphicx}
\usepackage{dcolumn}
\usepackage{bm}

\begin{document}

\title{Electronic structure of graphene functionalized with boron and nitrogen}

\author{ Magdalena Woi\'nska}
\affiliation{%
Faculty of Chemistry, University of Warsaw, ul. Pasteura 1,
PL-02-093 Warszawa, Poland
}
\author{Karolina Z. Milowska}
 \email{karolina.milowska@gmail.com}
\affiliation{%
 Institute of Theoretical Physics, Faculty of Physics, University of Warsaw, ul. Ho\.za 69,
PL-00-681 Warszawa, Poland
}
\author{Jacek A. Majewski}
\affiliation{%
 Institute of Theoretical Physics, Faculty of Physics, University of Warsaw, ul. Ho\.za 69,
PL-00-681 Warszawa, Poland
}

\date{\today}

\begin{abstract}We present a theoretical study of the structural and electronic properties of graphene monolayer functionalized with boron and nitrogen atoms substituting carbon atoms.  Our study is based on the ab initio calculations in the framework of the density functional theory. We calculate the binding energies of the functionalized systems, changes in the morphology caused by functionalization, and further the band gap energy  as a function of the concentration of dopants. Moreover, we address the problem of possible clustering of dopants at a given concentration. We define the clustering parameter to quantify the dependence of the properties of the functionalized systems on the distribution of B/N atoms.
 We show that clustering of B/N atoms in graphene is energetically unfavorable in comparison to the homogenous distribution of dopants. For most of the structures, we observe a nonzero energy gap that is only slightly dependent on the concentration of the substituent atoms.  
\end{abstract}


\maketitle   

\section{Introduction}

Graphene, a two-dimensional monolayer formed out of sp$^2$ hybridizated carbon atoms ordered in the hexagonal lattice, is a material with unique properties. Due to the hexagonal symmetry, its valence and conduction bands with linear dispersion cross at the so called K point and determine semimetallic character of graphene and its extremely high electron mobility. \cite{a1} This, in connection with excellent mechanical properties, renders graphene an ideal candidate for applications in flexible electronics.  However, the ability of generating controllable band gap is a prerequisite for effective applications of graphene in transistor based electronic devices. Therefore, an effective functionalization that would open the zero energy band gap in pristine graphene without significant deterioration of the remaining advantageous properties is searched for, and practically any thinkable way  of reaching this goal is currently investigated.

There exist three general methods of opening the band gap of graphene: application of external electric field, usage of graphene nanoribbons, and functionalization. One of the practical realizations of the third technique considered in this work is doping the graphene layer with boron and nitrogen \cite{a2,a13,a14,a15,a12}. These elements, standing nearby carbon in the periodic table, should act as electron acceptors (B) and donors (N), allowing for fabrication of contemporary devices in the manner of the CMOS silicon technology \cite{a16}. Here, we report new results of our theoretical studies of electronic properties of graphene doped with boron and nitrogen atoms.

\section{Calculation details}

Our studies of functionalized graphene layer are based on {\it ab initio} calculations within the framework of the spin polarized denstity functional theory (DFT) \cite{a3,a4}. Generalized Gradient Approximation of the exchange-correlation functional in Perdew-Burke-Ernzerhof (PBE) parametrization has been applied \cite{a5}. Calculations with periodic boundary conditions have been performed using Siesta package \cite{a6,a7}.  Valence electrons have been represented with double zeta basis sets of orbitals localized on atoms; with polarization functions also included. The influence of core electrons has been covered within pseudopotential formalism. Norm-conserving Troullier-Martins nonlocal pseudopotentials \cite{a8} used in this study have been cast into the Kleinman-Bylander \cite{a9} separable form. The energy cut-off determining the density of the utilized real space grid has been set to 800 Ry.The Brillouin zone has been sampled in the 5x5x1 Mokhorst-Park scheme. For the band structure and density of states calculations k-sampling changed to 15x15x1 scheme. Calculations were performed within the supercell geometry with graphene layers separated by a distance large enough to eliminate any interactions. Structural optimization has been conducted using the conjugate gradient algorithm to achieve residual forces acting on the atoms lower than 0.001 eV/$\AA$.

We have performed calculations for 5x5 supercells containing 25 graphene's primitive unit cells (i.e., consisting of 50 atoms). Such supercell has been chosen in order to examine wide range of dopant concentrations and a variety of possible distributions of the substituent atoms, and also to avoid band gap reduction caused by certain supercell symmetries \cite{a10}. In the described graphene supercell, one to ten B/N atoms have been introduced (leading to the corresponding concentration of 2-20$\%$). In the case of two substituent atoms in a supercell all the 11 symmetrically nonequivalent configurations of the atoms have been examined. In the remaining cases 12 different randomly chosen configurations have been taken into account. Optimized geometries and band structures, as well as related properties - energy band gaps, binding energies, and shifts of the Fermi level with respect to the top of the valence band and bottom of the conduction band for B and N dopants, respectively, have been also obtained for all the considered concentrations and symmetrically nonequivalent configurations of substituent atoms. All the values, for each concentrations, have been averaged over the investigated configurations and compared with relevant results for B/N substitution.

The binding energy per atom, which has been calculated according to the formula:
\begin{equation}
E_{b/n}  = \frac{1}{50 }\left( {E_{tot}  - \sum\limits_{\alpha  = 1}^{50} {E_{atom,\alpha } }  } \right),
\end{equation}
is used as the measure of the stability of the systems studied. $E_{tot}$ is total energy of functionalized graphene and $E_{atom,\alpha }$ is the total energy of the free atom of type $\alpha$ (i.e., C, N, or B).

To quantify the effect of B/N atoms distribution over the graphene lattice on the calculated properties of the doped systems, a special parameter measuring the level of clustering (further called clustering parameter) of atoms has been introduced. This parameter is, in general, linearly proportional to the sum of n-1 shortest nonequivalent distances between B/N atoms (where n is the number of B/N atoms in a supercell) and it is normalized to reach the value of -1 for maximal and 1 for minimal clustering in a system with geometry of pristine graphene. 
The described clustering parameter is given by the following mathematical formula:
\begin{equation}
\label{eq2}
c= \frac{{2\sum  - \sum _{\max }  - \sum _{\min } }}{{\sum _{\max }  - \sum _{\min } }},
\end{equation}
where $\sum$ is a sum of n-1 shortest nonequivalent distances between B or N atoms in the system measured after geometry optimization, $\sum_{max}$ and $\sum_{min}$  are maximal and minimal value that $\sum$ reaches for a given n for ideal hexagonal lattice.  We have assumed that both substituent atoms, B and N, create bonds among themselves one another equal or greater than C atoms in ideal hexagonal lattice and change the symmetry. 
Defined in this way, the clustering parameter for two dopants in the supercell (n=2) is the most negative, if B/N atoms are linked by a chemical bond, and the most positive if they are the furthest apart as possible for a given supercell. For example for the case of two N atoms, in considered 5x5 supercell, in which the the shortest distance in minimal clustering is 7.143 $\AA$ ($\sum$) after geometry optimization, the minimal and maximal distance in pristine graphene are equal to 1.433 $\AA$ ($\sum_{min}$) and 7.165 $\AA$ ($\sum_{max}$), respectively; the clustering parameter $c$ is 0.99.

\section{Results and discussion}

\begin{figure}[t]%
\includegraphics*[width=\linewidth]{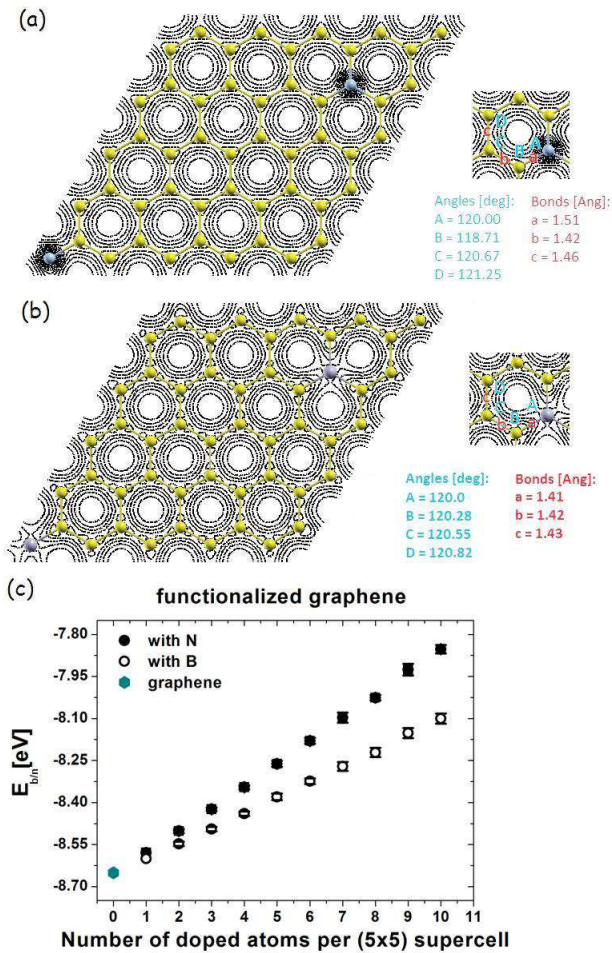}
\caption{%
The optimized structure and electron density depicted for clustering parameter close to 1 for two B/N atoms in 5x5 graphene supercell. Depletion of electron density in the vicinity of B atoms (a) and its higher values around the position of N atoms (b) can be observed. Bond lengths and angles between B/N and C atoms are also denoted for each case.  (c) The dependence of the maximum, minimum and averaged binding energy per atom (with standard deviations) for functionalized graphene on the concentration of B/N atoms. }
\label{struk}
\end{figure}

We start the presentation of results by considering  structural properties of  functionalized graphene monolayers. 
The geometries of optimized structures for two boron ($c$=0.97) and nitrogen ($c$=0.99) atoms in 5x5 graphene supercell, the cortesponding electron density, as well as bond lengths and angles between the substituent atom and carbon atoms are presented in Fig.~\ref{struk}(a) and (b). 

Boron atoms in comparison to nitrogen atoms have a greater influence on the geometry of the doped graphene layer, mostly due to the fact that the covalent radius of boron is larger than that of a carbon atom, whereas  nitrogen has the covalent radius similar to carbon. 
It turns out that the graphene monolayer functionalized with boron atoms is no longer flat because B atoms stick out from the surface. 
Similar observation was pointed out in Refs. \cite{a2,a13,a12}.

Analyzing isolines in  Fig.~\ref{struk}(a) and (b), one can see regions of electron density depletion around the positions of boron atoms, as well as an increase of electron density in the vicinity of nitrogen atoms. Substituent atoms influence only the electron density around those carbon atoms with which they are linked by a chemical bond. Electronic density around the second and further neighbors remains unchanged.

To measure stability of the considered structures, we have plotted, in Fig.~\ref{struk}(c), the dependence of the binding energy per atom on concentration of B/N  atoms. Both types of structures do not segregate after the geometry optimization and binding energies are more negative,  for B-functionalized graphene than for N-functionalized one. Therefore the boron doped structures are more stable and the differences in binding energies between B- and N-doped graphene increase with growing concentration of dopants, from 0.02 eV up to 0.25 eV for 2$\%$ and 20$\%$ concentration, respectively. Our observations are in agreement with previous theoretical work \cite{a17} which shows that substitution with boron atoms costs less energy than substitution with nitrogen atoms. Morover, boron doped monolayer graphene and nitrogen doped monolayer graphene were already sythesized by CVD method \cite{a13,a16,a18} and are stable enough for studying their physical and chemical properties.

\begin{figure}[t]%
\includegraphics*[width=\linewidth]{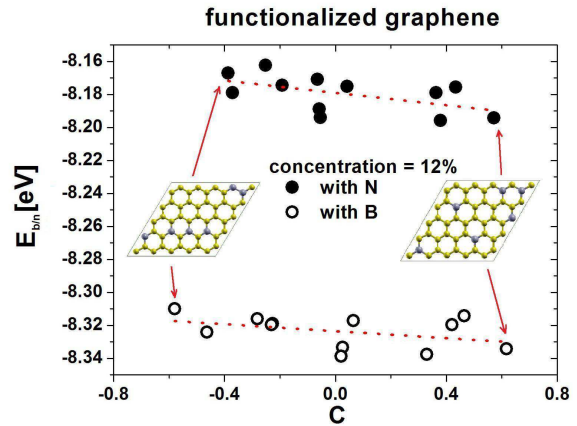}
\caption{%
The dependence of binding energy per atom on clustering parameter for 12$\%$ concentration of B and N atoms.}
\label{klaster}
\end{figure}

To account for the influence of the distribution of B/N atoms in a supercell on stability of the functionalized graphene, the clustering parameter $c$ has been calculated according to the formula given by Eq.~\ref{eq2} for each concentration of dopants. Exemplary results for 12$\%$ concentration both for boron and nitrogen, are shown in Fig.~\ref{klaster}. As one can see, from the  (dotted) lines fitted to each case, the binding energy barely dependends on the clustering level. For nitrogen, as well as for boron, the configuration of dopants being closer to its own species is less energetically preferable than being surrounded and separated by carbon atoms.  To have a better insight into this problem, further calculations including more symmetrically nonequivalent realizations of each concentration should be supportive.

\begin{figure}[t]%
\includegraphics*[width=\linewidth]{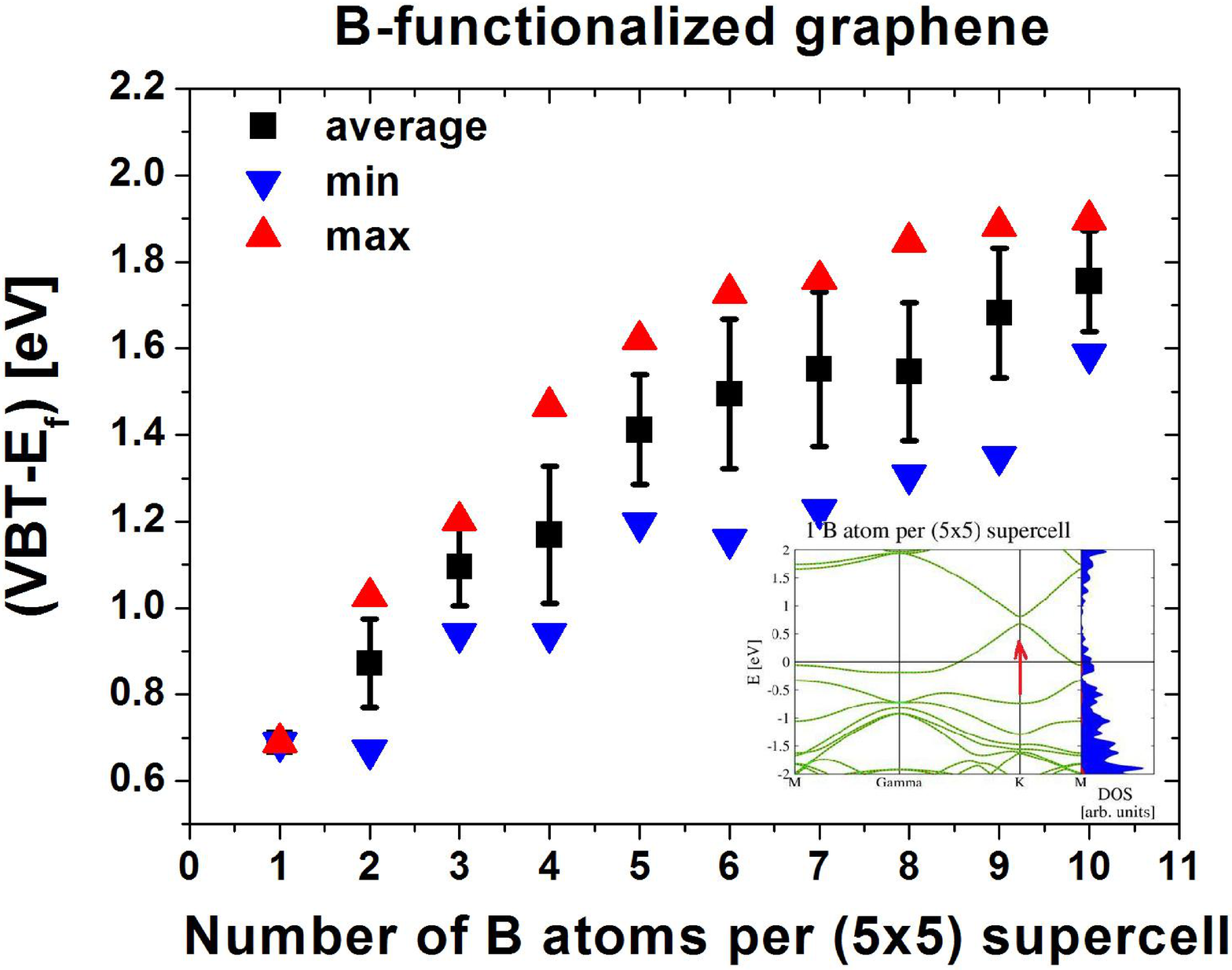}
\includegraphics*[width=\linewidth]{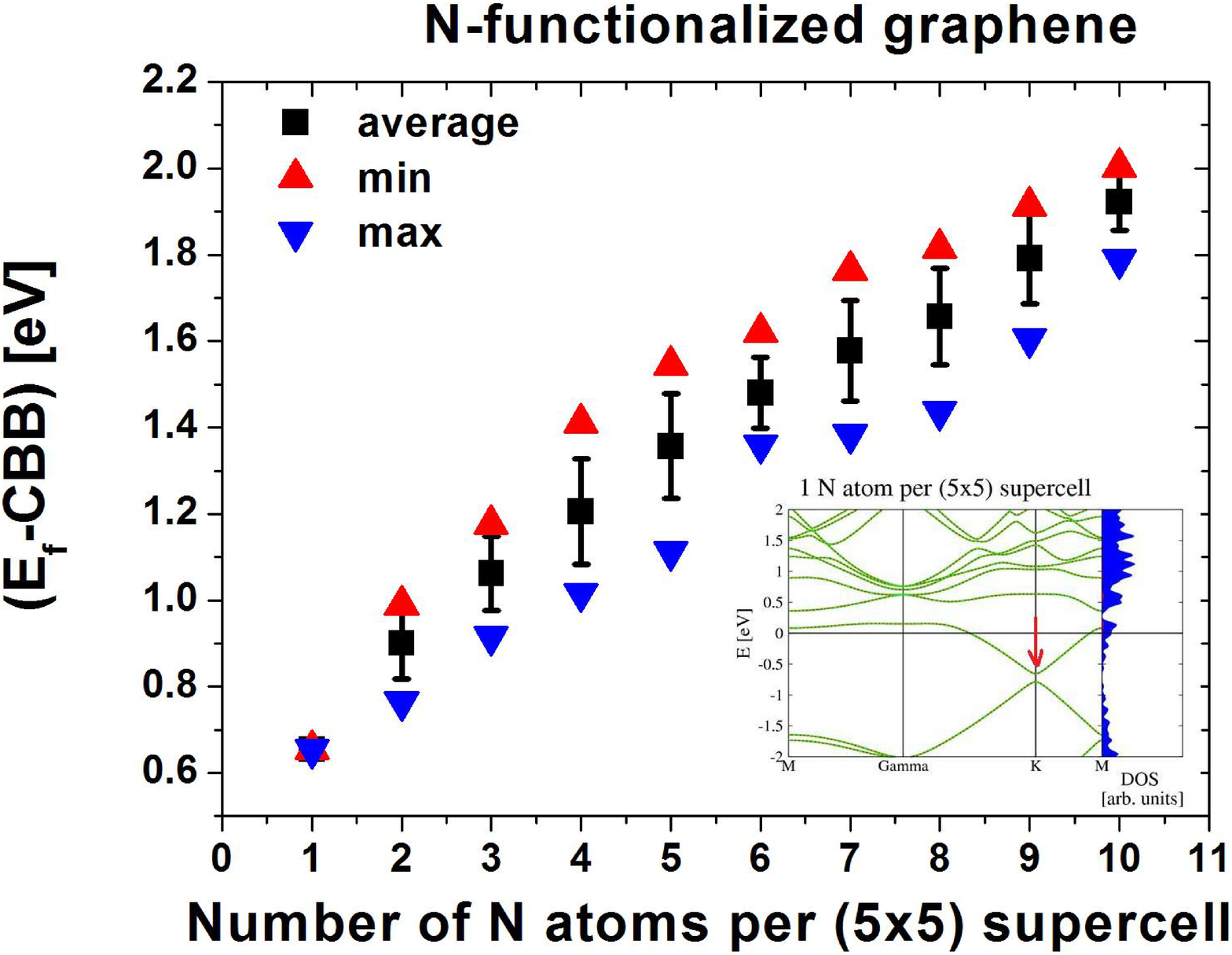}
\caption{%
The maximum, minimum and averaged  (with standard deviations) shift of the Fermi level from (a) the top of the valence band (VBT) for B and (b) the bottom of the conduction band (CBB) for N in function of concentration of substituent atoms. Inset: band structure and density of states for the case of one substituent atom in 5x5 supercell. Shifting of Fermi energy is marked with red arrows.}
\label{efermi}
\end{figure}

Fermi level shift with respect to VBT\footnote{Valence band top corresponds to HOMO level in band structure of pure graphene.}/CBB\footnote{Conduction band bottom  corresponds to LUMO level in band structure of pure graphene.} resulting from B/N functionalization and loss/gain of one electron per substituent atom is depicted in Fig.~\ref{efermi}. Similar observation, but for smaller concentrations was denoted in Refs. \cite{a14,a15,a11}.

\begin{figure}[t]%
\includegraphics*[width=\linewidth]{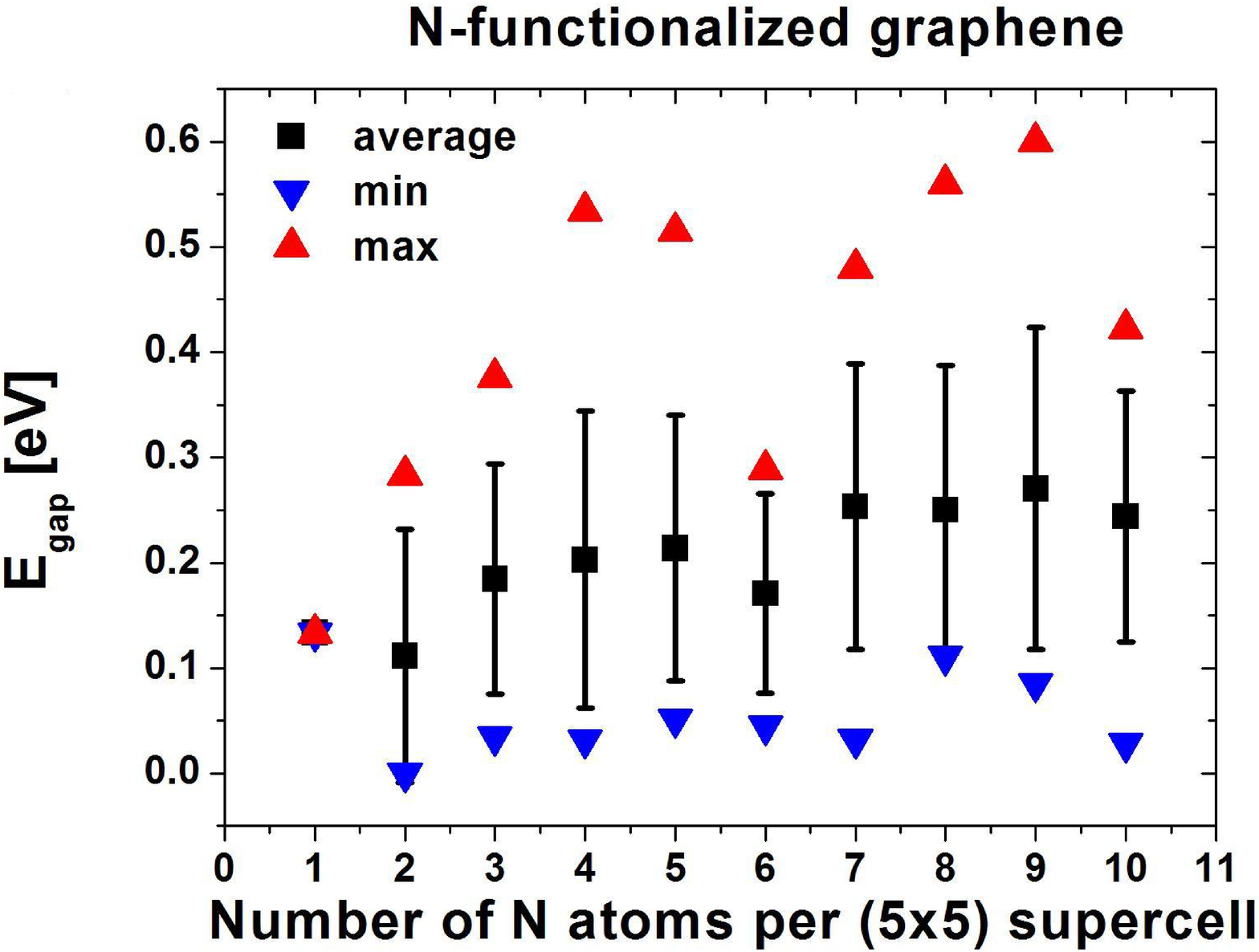}
\caption{%
The dependence of the maximum, minimum and averaged band gap (with standard deviation indicated) for the N-functionalized graphene on the concentration of N atoms.}
\label{egap}
\end{figure}

Within each of the considered concentrations of B/N atoms, we have performed electronic structure calculations for different configurations of dopants corresponding with this concentration. In the next step, we have performed averaging of the calculated band gaps. The average band gaps for N-doped graphene together with the maximal and minimal values of the band gaps are presented in Fig.~\ref{egap}. For B-doped graphene, the picture is analogous and is therefore not presented here. Opening of the band gap is observed in case of the majority of configurations, with maximal gap amounting to almost 0.5 eV for B- and 0.6 eV for N-doping. The mean value of the band gap slightly increases with the growing number of substituent atoms. This trend is more pronounced for lower concentrations, where the band gaps range from about 0.1 eV for two atoms per supercell to nearly 0.25 eV for 7-10 atoms. The average band gap saturates slightly for higher concentrations. In the case of six B/N atoms in the supercell, a decrease in an average energy gap width by over 0.5 eV can be noticed in comparison with five B/N atoms. It can indicate that averaging over 12 random configurations may not be sufficient to eliminate the strong influence of some particular configurations on the average values of the band gap. The results of our calculations signalize a limited possibility of  band gap engineering by means of increasing the concentration of functionalizing species.

\section{Conclusions}

In this paper we report the results of {\sl ab initio} calculations for graphene functionalized with boron and nitrogen atoms. We investigate structures with one to ten borons or nitrogens in a graphene supercell for various symmetrically nonequivalent configurations. All the structures are stable with slightly more negative binding energies in the case of B-doping. Doping with B atoms introduces more changes in the morphology of the functionalized structures than doping with N atoms. Both types of dopants prefer to be homogenously distributed over the lattice rather than clustered. We conclude that the effects of the band gap oppening in the boron and nitrogen functionalized graphene are qualitatively and quantitatively very similar. However, the magnitude of the band gap could be strongly dependent on a particular configuration of dopants. To clarify this issue, the role of disorder should be further studied. 

\section{Acknowledgement}
This work has been supported by the European Founds for Regional Development within the SICMAT Project (Contact No. UDA-POIG.01.03.01-14-155/09). This research was supported in part by PL-Grid Infrastructure. Some of the computations were performed using a cluster of ICM Interdisciplinary Centre for Mathematical and Computational Modelling (Grant No. G47-5 and G47-16), University of Warsaw.


\begin{thebibliography}{[1]}

\bibitem{a1}
A.\,K. Geim, K.\,S. Novoselov,
Nat. Mater.  \textbf{6}, 183 (2007).

\bibitem{a2}
L.\,S. Panchakarla, K.\,S. Subrahmanyam, S.\,K. Saha, A.~Govindaraj, H.\,R. Krishnamurthy, U.\,V. Waghmare, C.\,N.\,R. Rao, 
Adv. Mater. \textbf{21}, 4726 (2009).

\bibitem{a13}
T.~Wu, H.~Shen,L.~Sun,B.~Cheng, B.~Liua and J.~Shen,
New J. Chem.,\textbf{36} , 1385 (2012).

\bibitem{a14}
L.~Zhao, R.~He, K.\,T. Rim, T.~Schiros,K.\,S. Kim, H.~Zhou,Ch.~Gutiérrez,S.\,P. Chockalingam, C.\,J. Arguello, L.~Palova, D.~Nordlund, M.\,S. Hybertsen, D.\,R. Reichman, T.\,F. Heinz,P.~Kim, A.~Pinczuk, G.\,W. Flynn, A.\,N. Pasupath,
Science \textbf{333}, 999 (2011)

\bibitem{a15}
B.~Zheng, P.~Hermet and L.~Henrard,
ACS Nano, \textbf{4}, 4165 (2010).

\bibitem{a12}
S.~Gao, Z.~Ren., L.~Wan, J.~Zheng, P.~Guo, Y.~Zhou,
App. Surf. Science \textbf{257},7443 (2011). 

\bibitem{a16}
V.~Georgakilas~, M.~Otyepka, A.\,B. Bourlinos, V.~Chandra, N.~Kim, K.\,Ch. Kemp, P.~Hobza, R.~Zboril, K.\,S. Kim,
Chem. Rev. \textbf{112}, 6156 (2012)

\bibitem{a3}
P.~Hohenberg, W.~Kohn, 
Phys. Rev. \textbf{136}, 864 (1964).

\bibitem{a4}
W.~Kohn, L.\,J. Sham, 
Phys. Rev. \textbf{140}, A1133 (1965).

\bibitem{a5}
J.\,P. Perdew, K.~Burke, M.~Ernzerhof, 
Phys. Rev. Lett. \textbf{77}, 3865 (1996).

\bibitem{a6}
D.~Sanchez-Portal, P.~Ordejon, E.~Artacho, J.\,M. Soler, 
Int. J. Quantum Chem. \textbf{65}, 453 (1997).

\bibitem{a7}
J.\,M. Soler, E.~Artacho, J.~Gale, A.~Garcia, J.~Junquera, P.~Ordejon, D.~Sanchez-Portal, 
J. Phys.:Condens. Matter \textbf{14}, 2745 (2002).

\bibitem{a8}
N.~Troullier and J.\,L. Martins, 
Phys. Rev. B \textbf{43}, 1993 (1991).

\bibitem{a9}
L.~Klienman and D.\,M. Bylander, 
Phys. Rev. Lett. \textbf{48}, 1425 (1982).

\bibitem{a10}
J.\,M. García-Lastra,
Phys. Rev. B \textbf{82}, 235418 (2010).


\bibitem{a17}
N.~Berseneva, A.\,V. Krasheninnikov, R.\,M. Nieminen
PRL \textbf{107}, 035501 (2011).


\bibitem{a18}
L.\,S. Panchakarla, A.~Govindaraj, C.\,N.\,R. Rao
Inorganica Chimica Acta \textbf{363}, 4163 (2010).


\bibitem{a11}
E.~Beheshti, A.~Nojeh, P.~Servati,
Carbon \textbf{49}, 1561 (2011).



\end{thebibliography}
\end{document}